\documentclass[aps, pra, showpacs, twocolumn, amsfonts, amsmath, amssymb, superscriptaddress, floatfix,standalone]{revtex4}

\usepackage[T1]{fontenc}
\usepackage[latin9]{inputenc}
\usepackage{color}   
\usepackage{graphicx}
\usepackage{ulem}
\usepackage{float} 
\usepackage{fixltx2e}

\graphicspath{pics}

\def\be{\begin{equation}} 
\def\ee{\end{equation}}     
\def\bea{\begin{eqnarray}}      
\def\eea{\end{eqnarray}}

\begin{document}

\title{Emergence of dark soliton signatures in a one-dimensional unpolarized attractive Fermi gas on a ring}

\author{Andrzej Syrwid} 
\affiliation{
Instytut Fizyki imienia Mariana Smoluchowskiego, 
Uniwersytet Jagiello\'nski, ulica Profesora Stanis\l{}awa \L{}ojasiewicza 11 PL-30-348 Krak\'ow, Poland}

\author{Dominique Delande} 
\affiliation{ 
Laboratoire Kastler Brossel, Sorbonne Universit\'{e}s, CNRS,
ENS-Universit\'{e} PSL, Coll\`{e}ge de France, 4 Place Jussieu, 75005 Paris, France} 
 
\author{Krzysztof Sacha}  
\affiliation{
Instytut Fizyki imienia Mariana Smoluchowskiego, 
Uniwersytet Jagiello\'nski, ulica Profesora Stanis\l{}awa \L{}ojasiewicza 11 PL-30-348 Krak\'ow, Poland}
\affiliation{Mark Kac Complex Systems Research Center, Uniwersytet Jagiello\'nski, ulica Profesora Stanis\l{}awa \L{}ojasiewicza 11 PL-30-348 Krak\'ow, Poland
}


\begin{abstract}
The two-component Fermi gas with contact attractive interactions between different spin components can be described by the Yang-Gaudin model. Applying the Bethe ansatz approach, one finds analytical formulae for the system eigenstates that are uniquely parametrized by the solutions of the corresponding Bethe equations.  Recent numerical studies of the so-called {\it yrast} eigenstates, i.e. lowest energy eigenstates at a given non-zero total momentum, in the Yang-Gaudin model show that their spectrum resembles {\it yrast} dispersion relation of the Lieb-Liniger model which in turn matches the dark soliton dispersion relation obtained within the nonlinear Schr\"{o}dinger equation. It was shown that such conjecture in the case of the Lieb-Liniger model was not accidental and that dark soliton features emerged in the course of measurement of positions of particles, when the system was initially prepared in an {\it yrast} eigenstate. Here, we demonstrate that, starting with {\it yrast} eigenstates in the Yang-Gaudin model, the key soliton signatures are revealed by the measurement of pairs of fermions.  We study soliton signatures in a wide range of the interaction strength. 
\end{abstract} 

\maketitle

\section{Introduction}
\label{intro}

Non-linear wave equations can possess solitonic solutions that propagate without any change of their shape. These extraordinary structures appear in a wide range of physical systems and may be formed by electromagnetic waves in non-linear optics \cite{KivsharOpticalSol}, as well as matter waves particularly investigated in ultra-cold atomic gases. Bose-Einstein condensates (BEC), made up by a single-component Bose gas cooled to nearly  absolute zero temperature, turn out to be excellent playgrounds for the investigation of matter wave solitons \cite{burger1999,denschlag2000,strecker2002,khaykovich2002,becker,Stellmer2008,Weller2008,Theocharis2010,  Gawryluk2006, Pawlowski2015}. In the mean field description, we assume that every single atom in a BEC experiences an effective average potential and occupies exactly the same single particle state. Such a case is described by the Gross-Pitaevskii equation (GPE) that possesses bright and dark soliton solutions in one dimensional (1D) space for attractive and repulsive interparticle interactions, respectively \cite{pethicksmith}. The experimental realization of both kinds of solitons confirmed the theoretical predictions obtained within the GPE \cite{burger1999,denschlag2000,strecker2002,khaykovich2002,becker,Stellmer2008,Weller2008,Theocharis2010,Boisse2017}. 
The observation of the quantum nature of solitons, i.e. many-body effects that go beyond the mean field GPE description  \cite{Lai89,Lai89a,corney97,corney01, castinleshouches,Dziarmaga2002,Law2003,dziarmaga03,dziarmaga2004, dziarmaga06,Streltsov2008,Weiss09,sacha2009,Mishmash2009_1,Mishmash2009_2, Dziarmaga2010,Mishmash2010,martin2010b,Gertjerenken2012,Gertjerenken2013, delande2013,delande2014, kronke15,Hans2015, Syrwid2015,Syrwid2016,Marchant2016, Syrwid2017, Katsimiga2017a, Katsimiga2017b, Oldziejewski2018}, is still very challenging from an experimental point of view. Nevertheless, the rapid development of laboratory techniques devoted to investigations of ultra-cold atomic gases gives an opportunity to study systems dominated by quantum many-body effects  
\cite{
Boisse2017}.  

The experimental observation of solitons in Bose systems provoked the investigations of similar structures in Fermi systems.
The two-component Fermi gas with attractive interactions between fermions with different internal degrees of freedom can form a superfluid state which can be described by a set of non-linear Bogoliubov-de Gennes equations in the Baarden-Cooper-Schrieffer (BCS) regime \cite{pethicksmith}. Although this approach is dedicated to the determination of ground state properties, it can also be used to describe dark soliton solutions, where  particle densities are very similar to those of the ground state, but where the BCS pairing function reveals signatures of a dark soliton  \cite{DziarmagaSacha2004,DziarmagaSacha2005,Antezza2007}. Passing the BCS-BEC crossover, the BCS pairing function, with dark soliton signatures, becomes the dark soliton wavefunction of a molecular BEC \cite{Antezza2007,Pieri2003}.

The experimental realization of dark solitons in a BEC is based on the phase imprinting method \cite{burger1999,Dobrek1999,denschlag2000,Andrelczyk2001,Carr2001,becker,Stellmer2008}. The phase of the condensate can be modified by the application of a short laser pulse whose intensity varies over the atomic cloud. In particular, it is possible to carve a dark soliton notch so that a half of a condensate cloud acquires a $\pi$ phase. It turns out that, using the same phase engineering technique, one can observe the generation of pairs of dark and bright soliton-like states in noninteracting single-component fermionic system, where the Pauli blocking plays the role of interparticle repulsion \cite{Karpiuk2002a,Karpiuk2002b}.  The same idea was applied to create a dark soliton in a superfluid Fermi system \cite{Yefsah2013}. However, the resulting state quickly decayed to a vortex that has been displayed in numerical simulations \cite{Bulgac2014,Scherpelz2014} and observed in the subsequent experiment  \cite{Yefsah2014}. The analysis of the nature of the dark soliton BCS pairing function suggests that, in order to excite a dark soliton in a superfluid Fermi system, only one fermion of a Cooper pair has to undergo the phase imprinting procedure  \cite{SachaDelande2014}.   
 
In general, our understanding of quantum many-body systems is very limited. Fortunately, there are many-body systems in lower dimensions for which the brilliant  method of Bethe ansatz is applicable \cite{Bethe31}. This is exactly the case of one-dimensional nonrelativistic Bose and Fermi gases with particle interactions described by point-like contact potentials \cite{Korepin93, Gaudin, Oelkers2006}. In comparison to the case of identical bosons (Lieb-Liniger model \cite{Lieb63,Lieb63a}), the problem of the multi-component Fermi gas with contact interactions requires a generalization of the Bethe ansatz procedure. The model of a Fermi gas consisting of arbitrary numbers of fermions with two internal degrees of freedom has been solved analytically by Yang and Gaudin \cite{Gaudin67,Yang67}. Such an ultracold two-component Fermi gas has been the subject of extensive studies \cite{Fuchs2004,Tokatly2004,Mora2005,Iida2005,Batchelor2006,Orso2007,HuLiu2007,Guan2007,Chen2010,Guan2013,Shamailov2016}. 

The second branch of elementary excitations (the type II excitations) of the Lieb-Liniger model with periodic boundary conditions corresponds to the so-called \textit{yrast} states, referring to the lowest energy at a given non-zero total momentum. For weak repulsive interactions, the type II eigenstates have been associated with dark solitons due to the coincidence between  the \textit{yrast} spectrum of the Lieb-Liniger model and the dark soliton dispersion relation obtained within the mean field approach \cite{kulish76, ishikawa80}.
The conjecture was underpinned by other strong arguments presented in many publications
\cite{komineas02, jackson02, kanamoto08, kanamoto10, karpiuk12, karpiuk15,sato12, sato12a, sato16,Gawryluk2017}. 
The direct observation of the emergence of dark soliton signatures during the successive measurement of positions of particles, for the system initially prepared in the type II eigenstate, has been reported recently \cite{Syrwid2015,Syrwid2016}. A similar analysis led to the identification of dark 
soliton-like eigenstates of the Lieb-Liniger model in the presence of an infinite square well potential \cite{Syrwid2017}.  

For periodic boundary conditions, all energy eigenstates are invariant under translations of all particles by the same distance. Therefore, the reduced single particle density is uniform and cannot display any soliton signature. Such a feature of the type II eigenstates was the main impediment during the investigations of their soliton character. The unequivocal connection between dark solitons and \textit{yrast} excitations in the Lieb-Liniger model resulted in a broader examination of the solitonic nature of \textit{yrast} states. An ultracold balanced (unpolarized) gas of spin--$\frac{1}{2}$ fermions can be described within the Yang-Gaudin model \cite{Oelkers2006, Gaudin67,Yang67,Fuchs2004,Mora2005,Iida2005,Batchelor2006,Orso2007,HuLiu2007,Guan2007,Chen2010,Guan2013,Shamailov2016}.  In the presence of attractive interaction, two fermions with opposite spins tend to create a two-particle bound state. In the many-body case, there are two physically different regimes corresponding to weak and strong interaction limits. The first one refers to the BCS-like Cooper pair formation, for which the size of the pairs is larger than the mean pair separation. Tightly bound pairs can be observed in the second case when the attraction is very strong. The thermodynamic description reveals that the strongly attractive Yang-Gaudin model is closely related to a strongly interacting gas of bosonic dimers described by the Lieb-Liniger model. That is, the ground state energy of tightly bound pairs of fermions coincides with the energy of the attractive Bose gas, described by the Lieb-Liniger model, which forms a highly excited   super Tonks-Girardeau phase \cite{Guan2013,Shamailov2016,Chen2010,Astrakharchik2004,Astrakharchik2005,Batchelor2005,Haller2009}. The latter can be described by a system of attractive hard rods \cite{Chen2010}. In the limit of infinitely strong interactions, the energy of the super Tonks-Girardeau phase matches the ground state energy of the Tonks-Girardeau gas described by the Lieb-Liniger model of strongly repulsive bosons \cite{Chen2010}.
Note that the pairing phenomenon in similar systems confined in a harmonic trap was meticulously analyzed in Ref. \cite{SowinskiGajda2015}.

Although the link between the repulsive Lieb-Liniger gas and the attractive Fermi system described by the Yang-Gaudin Hamiltonian is not entirely understood, it is expected that the \textit{yrast} excitations of the Fermi gas in question may correspond to dark solitons. The supposition is additionally supported by recent results showing that the spectrum of \textit{yrast} excitations in the Yang-Gaudin model is very similar to the corresponding type II spectrum of the Lieb-Liniger model which, in turn, matches the dispersion relation of dark solitons in the weak interaction limit \cite{Shamailov2016}. In analogy with the Bose case, eigenstates of the Yang-Gaudin system are translationally invariant when we impose periodic boundary conditions  
\cite{Shamailov2016}. 
Hence, we may expect that dark soliton signatures are hidden in the translationally symmetric \textit{yrast} states and may be observed only by the analysis of higher order correlation functions.

The present paper is devoted to the analysis of attractively interacting unpolarized systems of spin--$\frac{1}{2}$ fermions with attractive contact interactions between different spin components, confined in a ring geometry. By applying the Bethe ansatz approach, we investigate the formation of pairs of $\downarrow$--$\uparrow$ fermions and determine their size in a wide range of interaction strength, when the system is prepared either in the ground state or in the \textit{yrast} state. As pointed out in \cite{Shamailov2016}, we observe the crossover between two significantly different physical regimes corresponding to a BCS-like gas and a gas of impenetrable bosonic dimers, when the relevant dimensionless interaction parameter $\gamma\approx -1$. Following the Monte Carlo method \cite{Syrwid2015,Syrwid2016,Syrwid2017} we repeatedly perform the successive measurement of positions of particles, revealing the dark soliton signatures, i.e. a density notch and a phase flip in the wave function of the last anticipated pair of $\downarrow$--$\uparrow$ fermions. In addition, we analyze how the increasing number of particles affects the soliton structure.

\section{Yang-Gaudin model}
\label{YangGaudinModel}

A nonrelativistic ultracold gas of spin--$\frac{1}{2}$ fermions with inter-components interactions given by  point-like $\delta$--function potential in 1D  can be  described by the Yang-Gaudin model \cite{Gaudin,Oelkers2006,Gaudin67,Yang67,Guan2013,Recher2013,Recher2013a}.  Assuming that we deal with a system at zero temperature containing  $N_\downarrow \geq N_\uparrow$ spin-down and spin-up particles of equal masses $m=m_\downarrow =m_\uparrow=\frac{1}{2}$, the Hamiltonian reads 
\be
\mathcal{H}=-\sum_{j=1}^{N_{\downarrow}}\frac{\partial^2}{\partial x_j^{\downarrow 2}}-\sum_{s=1}^{N_{\uparrow}}\frac{\partial^2}{\partial x_s^{\uparrow 2}} + 2c\sum_{j=1}^{N_\downarrow}\sum_{s=1}^{N_\uparrow} \delta(x_j^\downarrow -x_s^\uparrow),
\label{h}
\ee
where the units have been chosen such that  $\hbar=1$. The number of particles in  each single component $N_{\downarrow,\uparrow}$ is a conserved quantity.  Note that the particles belonging to different spin components can be distingushed because there is no spin flipping term in the Hamiltonian in Eq.~(\ref{h}).  The interaction strength is measured by the following dimensionless parameter  
\be
\gamma=\frac{c}{n},
\ee
with $n=\frac{N_\downarrow+N_\uparrow}{L}$ denoting the average particle density in the system of size $L$. 

While the Bethe ansatz formulation of the problem is very simple, i.e. the eigenstates are superpositions of plane waves, the structure of the resulting wave functions is very cumbersome \cite{Guan2007,Guan2013,Gaudin,Oelkers2006,Yang67,Gaudin67,Recher2013,Recher2013a} and hard to use in numerical calculations. Fortunately, solutions of the Yang-Gaudin model can be rewritten in the determinant form \cite{Recher2013,Recher2013a}
\be
\Psi(\{x^\downarrow\},\{x^\uparrow\},\{k\},\{\Lambda\}) \propto \sum_{\pi\in \mathcal{S}_{N_\uparrow}} \mathrm{sgn}(\pi) \,\mathcal{W}_{\pi,\uparrow} \, \det  \Phi ,
\label{GYSol}
\ee
where 
\be
\mathcal{W}_{\pi,\uparrow}= \prod_{j<l}^{N_\uparrow} \left[ i\left( \Lambda_{\pi(j)}-\Lambda_{\pi(l)} \right) + c \, \mathrm{sign}(x_l^\uparrow -x_j^\uparrow) \right],
\label{GYSolW}
\ee
and the $(N_\downarrow+N_\uparrow)\times (N_\downarrow+N_\uparrow)$ matrix $\Phi$, represented by two rectangular matrices separated by the vertical bar, reads  
\begin{flushleft}
$\displaystyle{
\Phi= \left( \left[ \prod_{s=1}^{N_\uparrow}\mathcal{A}_j(\Lambda_{\pi(s)},x_l^\downarrow-x_s^\uparrow)\mathrm{e}^{i k_j x_l^\downarrow}\right]
 \right|
 }
 $\end{flushleft}
\be
\,\,\,\,\,\,\,\, \left. 
\left[ \prod_{s\neq m}^{N_\uparrow}\mathcal{A}_j(\Lambda_{\pi(s)},x_m^\uparrow-x_s^\uparrow)\mathrm{e}^{i k_j x_m^\uparrow}\right]
 \right)_{\substack{j=1,\ldots, N_\uparrow+N_\downarrow\\
                  l=1,\ldots, N_\downarrow\\
                  m=1,\ldots, N_\uparrow}},
\label{GYSolPhi}
\ee
with
\be
\mathcal{A}_j(\Lambda,x)=i(k_j-\Lambda)+\frac{c}{2} \, \mathrm{sign}(x).
\label{GYSolA}
\ee
The summation in Eq.~(\ref{GYSol}) is taken over all permutations $\pi$ of the permutation group $\mathcal{S}_{N_\uparrow}$. The parity of the permutation $\pi$ is extracted by $\mathrm{sgn}(\pi)=\pm 1$, while, for real $x$, the function $\mathrm{sign}(x)=x/|x|$. The eigenstates given by Eq.~(\ref{GYSol}) are uniquely parametrized by the sets of quasi-momenta $\{k_j\}_{j=1,\ldots,N_\downarrow + N_\uparrow}$ and spin-roots $\{\Lambda_s\}_{s=1,\ldots,N_\uparrow}$. The latter quantities are auxiliary and appear due to the existence of two internal degrees of freedom interpreted as opposite spin directions.  Since the Hamiltonian in Eq.~(\ref{h}) commutes with the total momentum operator 
\be
\mathcal{P}=-i \sum_{j=1}^{N_\downarrow}\frac{\partial}{\partial x_j^\downarrow}-i \sum_{s=1}^{N_\uparrow}\frac{\partial}{\partial x_s^\uparrow},
\label{GYTotMom}
\ee
the states $\Psi$  simultaneously satisfy the two eigenequations
\be
\mathcal{H}\Psi_{\{k\}}=E_{\{k\}}\Psi_{\{k\}}, \,\,\,\,\,\,\,\, \mathcal{P}\Psi_{\{k\}}=P_{\{k\}}\Psi_{\{k\}},
\label{GYEigEqs}
\ee
where the eigenvalues $E_{\{k\}}$ and $P_{\{k\}}$ are simply given by the quasi-momenta $k_j$ \cite{Recher2013,Recher2013a}
\be
E_{\{k\}}=\sum_{j=1}^{N_\downarrow + N_\uparrow}k_j^2, \,\,\,\,\,\,\,\, P_{\{k\}}=\sum_{j=1}^{N_\downarrow + N_\uparrow}k_j.
\label{GYEigs}
\ee
The symmetry properties of the wave functions $\Psi$,  
\be
\Psi(\rho_{\sigma}\{x^{\sigma}\},\{k\},\{\Lambda\})=\mathrm{sgn}(\rho_\sigma)\Psi(\{x^{\sigma}\},\{k\},\{\Lambda\}),
\label{GYSym1}
\ee
\be
\Psi(\{x^{\sigma}\},\tau\{k\},\{\Lambda\})=\mathrm{sgn}(\tau)\Psi(\{x^{\sigma}\},\{k\},\{\Lambda\}),
\label{GYSym2}
\ee
\be
\Psi(\{x^{\sigma}\},\{k\},\eta\{\Lambda\})=\mathrm{sgn}(\eta)\Psi(\{x^{\sigma}\},\{k\},\{\Lambda\}),
\label{GYSym3}
\ee
for arbitrary permutations $\rho_{\sigma=\downarrow,\uparrow} \in \mathcal{S}_{N_{\downarrow,\uparrow}}$, $\tau \in \mathcal{S}_{N_\downarrow}$ and $\eta \in \mathcal{S}_{N_\uparrow}$, are discussed in details in Refs. \cite{Recher2013,Recher2013a}.

Imposing periodic boundary conditions, i.e.
\be
\underset{j=1,\ldots,N_{\downarrow,\uparrow}}{\forall}\!\!\!: \,\,\, \Psi(\ldots, x_j^{\downarrow,\uparrow}+L, \ldots)=\Psi(\ldots, x_j^{\downarrow,\uparrow}, \ldots),
\label{GYPBC}
\ee
we obtain the following set of the so-called Bethe ansatz equations for the quasi-momenta $\{k\}$ and the spin-roots $\{\Lambda\}$ \cite{Yang67,Gaudin67,Guan2007,Guan2013,Oelkers2006,Shamailov2016,Batchelor2006} 
\be
\mathrm{exp}\left(i k_j L\right) =\prod_{n=1}^{N_\uparrow}\frac{k_j-\Lambda_n+i\frac{c}{2}}{k_j-\Lambda_n-i\frac{c}{2}}\Bigg|_{j=1,\ldots,N_\downarrow + N_\uparrow},
\label{GYPBCeqs1}
\ee
\be
\!\prod_{j=1}^{N_\downarrow+N_\uparrow}\!\frac{\Lambda_m\! -k_j\! +i \frac{c}{2} }{\Lambda_m\! -k_j\! -i \frac{c}{2}} =\!\prod_{\substack{n=1\\
                  n\neq m}}^{N_\uparrow} \!\frac{\Lambda_m\!-\Lambda_n\!+i c}{\Lambda_m\!-\Lambda_n\!-i c}\Bigg|_{m=1,\ldots, N_\uparrow}.
\label{GYPBCeqs2}
\ee

\section{Numerical method}
\label{NumMet}  

Despite the fact that the many-body eigenstates $\Psi$ can be cast into a superposition of determinants, the analysis of their properties is very burdensome. In fact, it is intractable to extract valuable physical information from every single determinant of the $\Phi$ matrix. Moreover, the number of terms in the summation that is present in Eq.~(\ref{GYSol}) dramatically proliferates with $N_\uparrow$. In order to investigate the features of $\Psi,$ we should examine the corresponding correlation functions. For this purpose, we have decided to numerically simulate the measurement of particle positions positions of particles.

In general, the above mentioned simulations are based on a one-by-one process of particle detection. Such an approach requires the calculation of conditional probability densities for measurements of consecutive particles \cite{javanainen96, dziarmaga03, dziarmaga06, Syrwid2015, Dagnino09, Kasevich20016}. Numerically, this method is extremely expensive in the considered system. The result of the measurement of $M=N_\downarrow+N_\uparrow$ particles can also be obtained by another method, i.e. by a direct sampling of the corresponding $M$-particle probability density employing the Monte Carlo algorithm of Metropolis {\it et al}. \cite{Metropolis1953}. By using the analytical expression for $M$-particle probability distribution $|\Psi(r_1,\ldots,r_M)|^2$ and following Refs.~\cite{Syrwid2016, Syrwid2017, Gajda_PauliCrystal}, we perform a so-called Markovian walk in the configuration space, generating a sequence of samples $\mathcal{R}=\{r_1,\ldots,r_M\}$ called a Markov chain. In our case, we assume that $r_{j}=x^\downarrow_j$ for $j=1,\ldots,N_\downarrow$ and $r_{N_\downarrow+j}=x^\uparrow_{j}$ for $j=1,\ldots,N_\uparrow$. Technically speaking, if $\mathcal{R}$ is the last element of the Markov chain, the next randomly chosen set of positions of particles $\mathcal{R}'$ is accepted with probability  $p=\mathrm{min}(1,|\Psi(\mathcal{R}')|^2/|\Psi(\mathcal{R})|)$. If $\mathcal{R}'$ is not accepted, we again append the set $\mathcal{R}$ at the end of the Markov chain.

The Metropolis procedure increases significantly the numerical efficiency but still allows for the studies of few body systems only. In order to investigate system containing more particles, one can employ numerical diagonalization of the Hamiltonian given by Eq.~(\ref{h}) in a truncated Hilbert space. Eigenstates of the system can be represented in the Fock state basis $\left[\prod_{j=s_{min}}^{s_{max}}\left| m_{j}^{\downarrow} \right>\right]\left[\prod_{j=s_{min}}^{s_{max}}\left| m_{j}^{\uparrow} \right>\right]$, where the $j$-th single particle mode $\phi_j(x)=L^{-1/2}\mathrm{exp}[i 2\pi j x/L]$ is occupied by $m_{j}^{\downarrow,\uparrow}=0,1$ particles. The numbers $s_{min}$ and $s_{max}$ determine the modes taken into account and have to be adjusted to reproduce the examined eigenstate accurately. The Hamiltonian in Eq.~(\ref{h}) commutes with $\mathcal{P}$, Eq.~(\ref{GYTotMom}), so in the chosen basis $\mathcal{H}$ is partitioned into blocks referring to different values of the total momentum $P$. Note that the following constraints have to be satisfied
\be
\sum_{j=s_{min}}^{s_{max}}\!\!\!m_j^\sigma=N_\sigma, \,\,\,\, P=\!\!\!\!\sum_{\sigma=\{\downarrow,\uparrow\}}\sum_{j=s_{min}}^{s_{max}}\!\!\!\frac{2\pi}{L}jm_{j}^{\sigma}=\frac{2\pi}{L}\mathcal{J},
\label{NumConstraints}
\ee
where $\mathcal{J}\in\mathbb{Z}$. 
 By definition the {\it yrast} states correspond to the lowest energy eigenvalue for a given total momentum (in fact, given by $\mathcal{J}$). 

The numerical diagonalization is used at the end of Sec.~\ref{Yrast} where influence of the total particle number on the soliton structures is considered.

\section{Attractive interactions: the ground state}
\label{GS}

 We start our considerations with the ground state in the presence of attractive interactions between different spin components ($c<0$).  Additionally, we restrict the discussion to the unpolarized system for which $N_\downarrow = N_\uparrow = N<\infty$. It has been shown that, in such a case, fermions with opposite spins tend to form  bound state pairs what is reflected by the appearance of conjugate pairs of quasi-momementa $k_{j,\pm}=\kappa_j \pm i \mu_j$, where $\kappa_j=\Re(k_{j,\pm}), \, \mu_j=|\Im(k_{j,\pm})|$ (see for example \cite{Oelkers2006,Batchelor2006,Guan2007,Guan2013}). The ground state solutions of the Bethe equations (\ref{GYPBCeqs1}) and (\ref{GYPBCeqs2}) in the weakly ($c\rightarrow 0_{-}$) and strongly ($c\rightarrow - \infty$) interacting limits are schematically depicted in Fig. 1.

\begin{figure}[h] 	        
\includegraphics[width=1.\columnwidth]{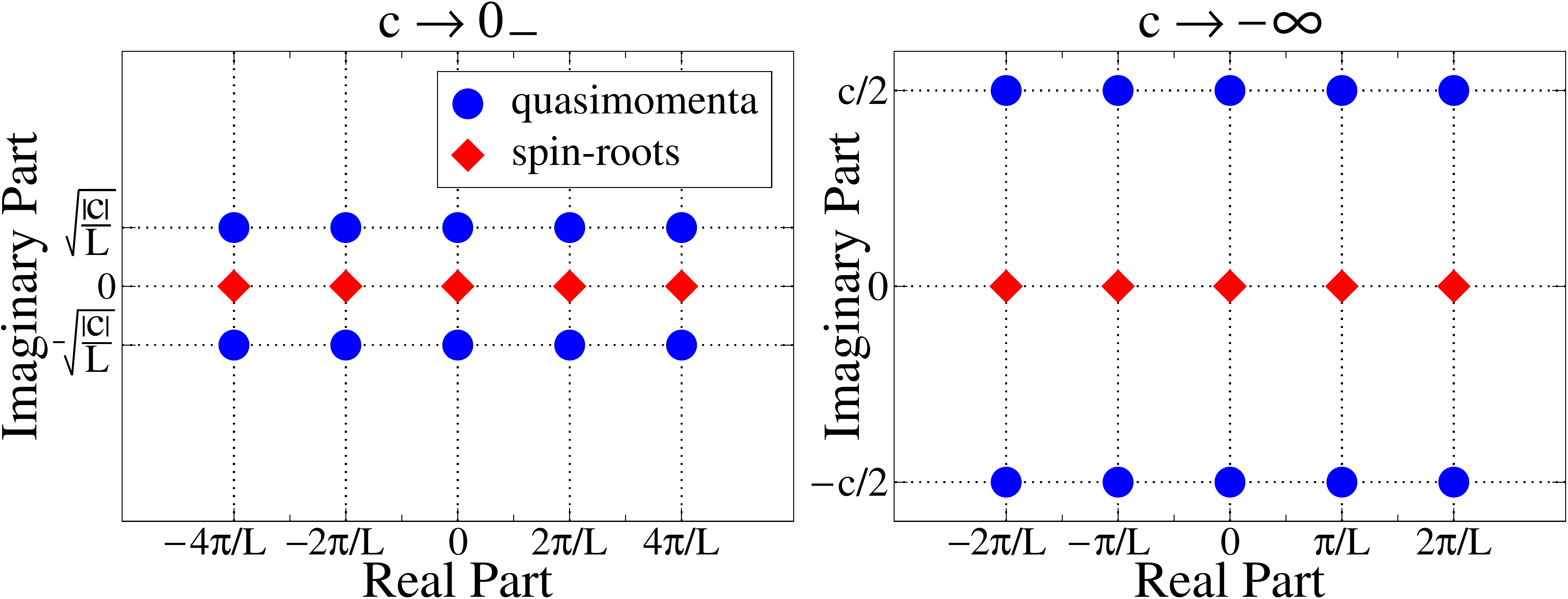}         
\caption{Sketch of the solutions of the Bethe equations (\ref{GYPBCeqs1}) and (\ref{GYPBCeqs2}) in the complex plane, for the balanced ground state with $N_\downarrow = N_\uparrow =N= 5$. Left (Right) panel corresponds to the regime of weak (strong) attraction $c\rightarrow 0_-$($c\rightarrow -\infty$). The  resulting quasi-momenta form conjugate pairs $k_{j,\pm}=\kappa_j\pm i \mu_j$. In both panels, the spin-roots $\Lambda_j$ are equal to $\kappa_j$.  All units are dimensionless.  }
\label{f1}
\end{figure}      

The expansion of the Bethe equations leads to the observation that in the two regimes in question, the ground state solutions take the following forms
\be
\begin{array}{lll}
\displaystyle{\lim_{\substack{c\rightarrow 0_-
}} \kappa_j=\lim_{\substack{c\rightarrow 0_-
}} \Lambda_j= \frac{2\pi}{L}n_j}, & &
                
  \displaystyle{  \mu_j  \stackrel{c\rightarrow 0_-}{\approx} \sqrt{\frac{|c|}{L}}}     ,
                \\ \\
\displaystyle{\lim_{\substack{c\rightarrow -\infty
}} \kappa_j=\lim_{\substack{c\rightarrow -\infty
}} \Lambda_j= \frac{\pi}{L}n_j}, & &
                
              \displaystyle{  \mu_j \stackrel{c\rightarrow -\infty}{\approx} \frac{|c|}{2}}     ,          
\end{array}
\label{Gs2regimes}
\ee
with $n_j\in \left\{-\frac{N-1}{2},\ldots, \frac{N-3}{2},\frac{N-1}{2}\right\}$ \cite{Oelkers2006,Guan2013}. The solutions can be interpreted as filling a "Fermi sphere" with the "Fermi surface" referring to the "Fermi momentum" $\pm \mathrm{max}_j(\Lambda_j)$. Note that the corresponding binding energies per $\downarrow$--$\uparrow$ pair, defined as $\varepsilon_B=  - \frac{1}{N}\sum_j[\Im(k_j)]^2=  - \frac{1}{N}\sum_j\mu_j^2$, are the following 
\be
\begin{array}{lll}
\displaystyle{ \varepsilon_B \stackrel{c\rightarrow 0_-}{\approx} -2\frac{|c|}{L}}  , & \,\,\,\,\,\, &
          
           \displaystyle{   \varepsilon_B    \stackrel{c\rightarrow -\infty}{\approx} -\frac{c^2}{2}}. 
\end{array}
\label{GsEnMom}
\ee

\begin{figure}[h] 	        
\includegraphics[width=1.\columnwidth]{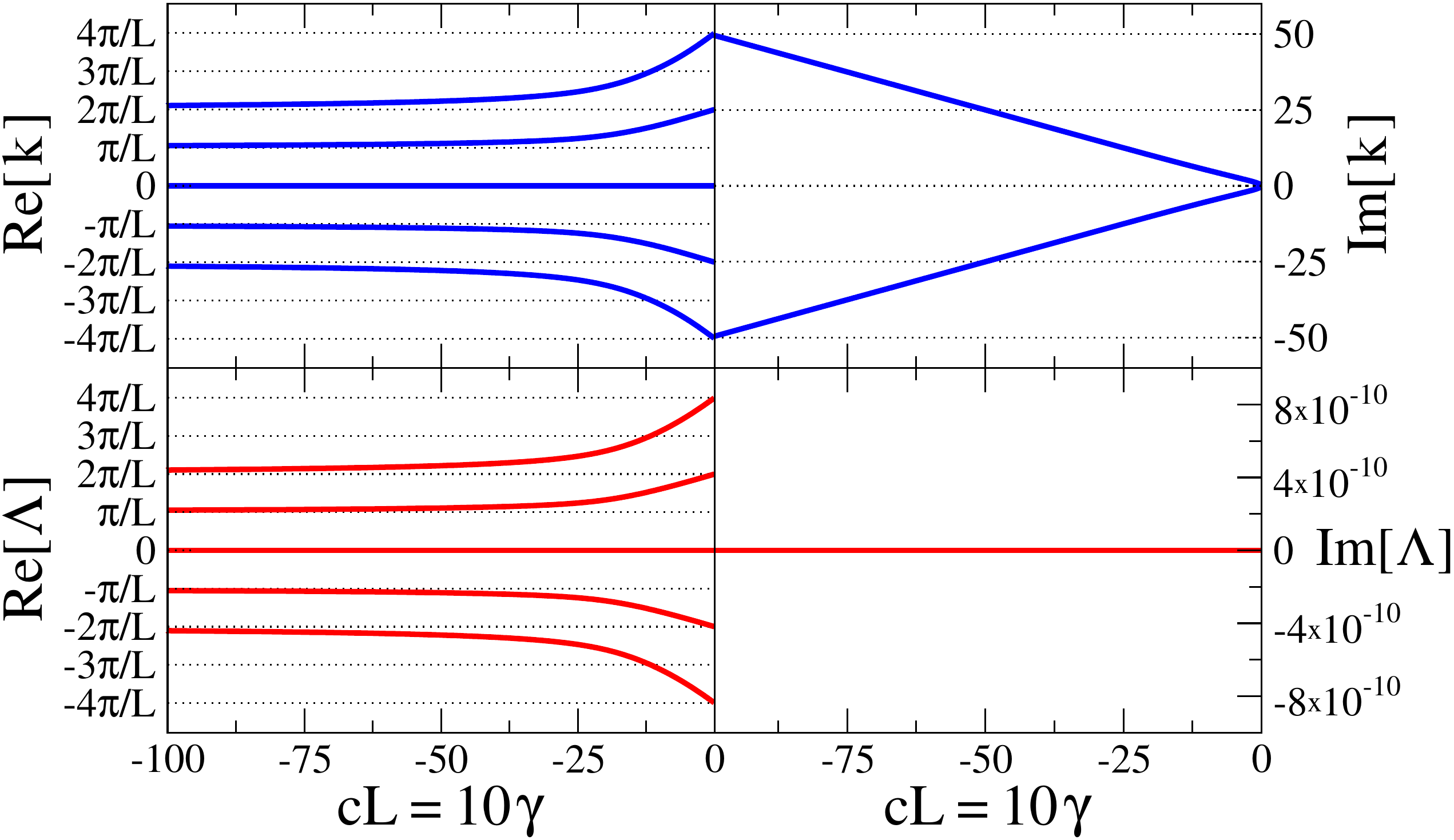}         
\caption{Solutions of the Bethe equations (\ref{GYPBCeqs1}) and (\ref{GYPBCeqs2}) for the ground state with $N_\uparrow=N_\downarrow= N=5$ versus the coupling constant $c$ ($\gamma=cL/2N$). While upper panels represent the real and imaginary parts of the resulting quasi-momenta $k_j$, lower panels refer to the spin-roots $\Lambda_j$ solutions. Note that, in the two limiting cases of weak and strong interaction strength, both quasi-momenta and spin-roots follow the predictions given by the Eqs.~(\ref{Gs2regimes}) depicted in Fig.~\ref{f1}. All units are dimensionless.}
\label{f2}   
\end{figure}

 In general, the Bethe equations are very difficult to solve for arbitrary values of  $c<0$. Therefore, it is convenient to start with one of the considered limits and, by employing a simple linear approximation, consecutively increase or decrease the coupling strength \cite{Shamailov2016}. The results of this procedure applied to the 5+5 particle ground state ($N=5$) are presented in Fig.~\ref{f2}. In the strongly attractive case, we can apply an additional approximation and simplify the Bethe equations.  That is, one can replace  $k_{j,\pm}$ by $\Lambda_j \pm i\frac{c}{2}$ and by simple algebraic manipulations obtain \cite{Shamailov2016,Sutherland2004}
\be 
2\Lambda_m L = 2\pi l_m-2\sum_{n=1}^{N}\mathrm{arctan}\!\left(\frac{\Lambda_m\!-\Lambda_n}{c}\right)\!\Bigg|_{m=1,\ldots,N}.
\label{GYString1}
\ee
where the soultions are determined by distinct  numbers $l_m$. By substituting $l_m=n_m$, cf. Eq.~(\ref{Gs2regimes}), we get the parametrization of the ground state of the balanced gas of fermions. 

\begin{figure}[h] 	        
\includegraphics[width=1.\columnwidth]{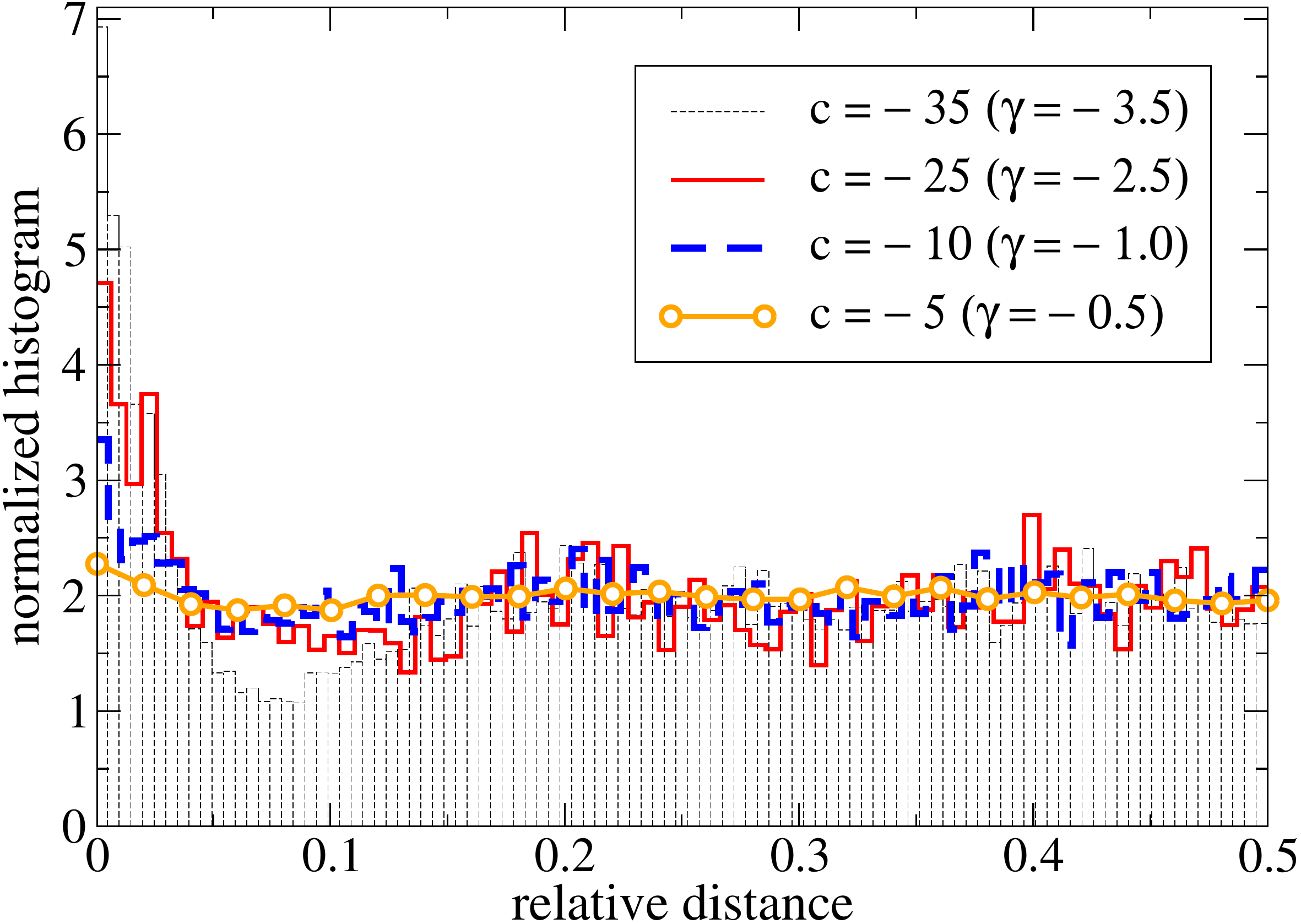}         
\caption{Histograms of the relative distance $\Delta_j^n$ [see Eq. (\ref{RelDist})] between fermions with opposite spins for different values of the coupling constant $c$. The considered 5+5 particles system of size $L=1$ was prepared initially in the ground state. The measurement of the positions of particles was performed with the help of the Metropolis algorithm, as described in Sec.~\ref{NumMet}. Increasing the inter-components attraction above $c=-10$ ($\gamma=c/2N=-1$) one observes an escalation of small-sized $\downarrow$--$\uparrow$ pairs occurrence. That is, when $\gamma<-1,$ the effective attraction causes the creation of dimers of size smaller than their mean separation. The background density $\approx 2$ dominates in all cases for distances larger than the average distance between particles belonging to the same spin component $\bar{\delta}=L/N=0.2$. The histograms have been prepared from the data collected from several millions of measurement realizations. All units are dimensionless.}
\label{f3}   
\end{figure}

Let us now analyze in details the problem of pairing of fermions belonging to different components. For this purpose, we have decided to investigate histograms of the relative distance between particles that are obtained in many measurement realizations. Dealing with the $N_\downarrow=N_\uparrow=N=5$ system and employing the Bethe ansatz solution, we perform numerical simulations of the particle measurement process with the help of the Metropolis routine (see Sec. \ref{NumMet}). In every single $j$-th realization of the detection procedure, we collect two sets of positions of particles i.e. $X_j^{\sigma}=\left\{x_{j,1}^\sigma,x_{j,2}^\sigma, \ldots, x_{j,5}^\sigma \right\}$ with $\sigma =\downarrow,\uparrow$. The relative distance on a ring of size $L$ can be defined as follows 
 \be 
\Delta_{j}^{n}=\mathrm{min}\left(\left|x_{j,n}^{\downarrow}-x_{j,n}^{\uparrow}\right|, \left|L-\left|x_{j,n}^{\downarrow}-x_{j,n}^{\uparrow}\right|\,\right| \right),
\label{RelDist}
\ee      
where the $j$ and $n$ indices refer to the measurement realization and to the particle number, respectively. By collecting all $\Delta_{j}^{n}$ distances (for all $j$ and $n$) after many realizations of the particle detection process, one can prepare a histogram of the relative distances between spin-up and spin-down fermions. In Fig.~\ref{f3}, we compare such histograms for different strengths of the inter-components attraction. We stress that, if one collects many measurement realizations, it does not matter which positions are taken in order to form pairs and calculate the distances as in Eq.~(\ref{RelDist}). In other words, the obtained histograms will not change if we randomly permute particles. This is the reason why the background density $\approx2$ appear in the histograms independently on the attraction strength (see Fig.~\ref{f3}).

The average distance between particles possessing the same spin $\bar{\delta}=L/N$ ($\bar{\delta}=0.2$ for  $L=1$ and $N=5$) is a reference quantity. When $\gamma = cL/2N \lesssim-1$ the $\downarrow$--$\uparrow$ pairing becomes visible in Fig.~\ref{f3}. That is, if the attraction is strong enough, the size of the pairs is smaller than $\bar{\delta}$. In this way we enter the regime of tightly bound pairs of fermions with opposite spins. On the other hand, we expect that formation of the Cooper-like pairs of size greater than $\bar{\delta}$ when $\gamma\gtrsim-1$ \cite{Shamailov2016}. 
A careful analysis of Fig.~\ref{f3} reveals oscillations in the profiles of the distributions with period  $\approx\bar{\delta}$. Moreover, in the strongly attractive case $c=-35$ $(\gamma=-3.5)$, one notices a density dip near the relative distance $0.08$. It is clear that, in such a regime, fermions are tightly bound and we deal with a gas of impenetrable bosonic dimers. The particles coming from the same spin component feel the Pauli exclusion. Hence, the $\downarrow$--$\uparrow$ molecules tend to distribute themselves uniformly in space. This simple mechanism is responsible for the oscillating behavior visible in Fig.~\ref{f3}. The same features can be observed within the BCS approach (see Ref.~\cite{KetterleZwirlein2008}) if we analyze the following correlation function  $\big< \hat{\psi}_\downarrow^\dagger (x) \hat{\psi}_\uparrow^\dagger (y)  \hat{\psi}_\uparrow (y)   \hat{\psi}_\downarrow (x)\big>$, where $\hat{\psi}_\sigma(x)$ are the canonical field operators and the average $\left<.\right>$ is taken in the BCS ground state. 

The above discussion concerning the creation of tightly bound molecules when $\gamma \lesssim-1$ stays in a very good agreement with the results which can be obtained for the two-body problem. The relative distribution of two distinguishable particles interacting via a contact attractive potential is well known, \cite{McGuire1964,Guan2013,CasitnHerzog2001}
 \be 
|\psi(r)|^2\propto\mathrm{e}^{ -|c| r },
\label{ProbDens2Body}
\ee
where  $r=\big|x^\downarrow-x^\uparrow\big|$ is the relative distance of the two particles. By rewriting $c=  2\gamma N/L=2\gamma/\bar{\delta},$ one obtains $|\psi(r)|^2 \propto\mathrm{exp}\big(-2|\gamma|  r/\bar{\delta} \big)$. Then the molecule size is given by  $\bar{\delta}/|\gamma| $. We will see that this simple two-body result matches the results obtained in the many-body simulations. Furthermore, it is now straightforward that the state in Eq.~(\ref{ProbDens2Body}) fits in the system of size $L=N\bar{\delta}$ for $\gamma \lesssim-1$.      

\begin{figure}[h] 	        
\includegraphics[width=1.\columnwidth]{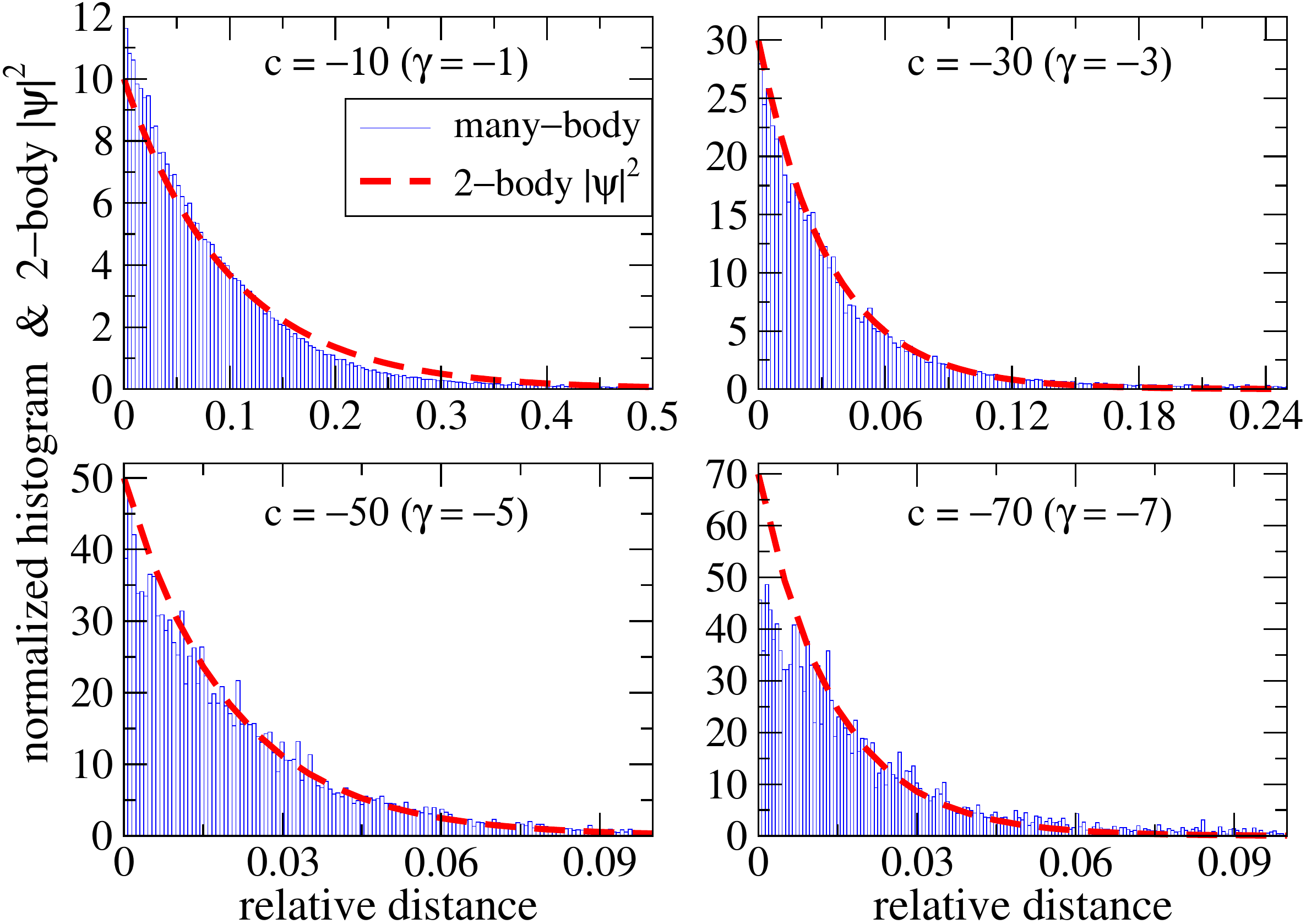}          
\caption{Distributions of  the relative distance between paired fermions belonging to different spin components. Normalized histograms represent the results obtained from many-body simulations that are based on the $\widetilde{\Delta}_j^n$ quantities defined in Eq.~(\ref{PairsRelDist}). Red dashed lines depict the corresponding two-body probability densities given by Eq.~(\ref{ProbDens2Body}). Note that the many-body results follow quite well the simple two-body solutions indicating domination of the two-body physics for $\gamma \lesssim -1$.  The system was prepared initially in the ground state. The size of the 1D space is $L=1$ and the number of particles $N_\downarrow=N_\uparrow=5$. All units are dimensionless. }  
\label{f4}       
\end{figure}                        

Armed with this knowledge, we can investigate the pairing phenomenon in the strongly attractive regime (when $\gamma\lesssim -1$) in details. In order to check how the pairing depends on the attraction strength, we have decided to determine which fermions belonging to the measurement collections $\{X_j^\downarrow\}$ and $\{X^\uparrow_j\}$ collections are actually paired. For this purpose, for every single $j$-th realization of the measurement process, we found the permutation $\tau\in \mathcal{S}_{N_\uparrow}$ minimizing the sum $\sum_n\widetilde{\Delta}_j^n$, where $\widetilde{\Delta}_j^n$ is defined as 
  \be 
\widetilde{\Delta}_{j}^{n}=\mathrm{min}\left(\left|x_{j,n}^{\downarrow}\!-x_{j,\tau(n)}^{\uparrow}\right|, \left|L\!-\left|x_{j,n}^{\downarrow}\!-x_{j,\tau(n)}^{\uparrow}\right|\,\right| \right),
\label{PairsRelDist}
\ee
and measures the relative distance between paired fermions. As before, we collect all the distances $\widetilde{\Delta}_j^n$ and prepare histograms corresponding to the distributions of the relative distances between fermions with opposite spins that are paired. It turns out that the obtained results match quite well the normalized 2-particle solutions (see Eq.~(\ref{ProbDens2Body})). Such an agreement confirms that the system is dominated by two-body physics for $\gamma\lesssim -1$. The many-body numerical outcomes and the above mentioned two-body results are presented in Fig.~\ref{f4}. We expect that the result holds true even for large $N$ because $\gamma\lesssim -1$ guarantees that the size of the $\downarrow$--$\uparrow$ molecules is smaller than $\bar{\delta}$. Note that the same analysis in the presence of weak attraction cannot be performed. That is, when the size of the anticipated $\downarrow$--$\uparrow$ pairs is larger than the average dimer separation $\bar{\delta},$ we do not have any practical tool to determine which fermions in the considered gas are paired.

\section{Attractive case: Yrast excitation}
\label{YrastPair}

Let us now consider an {\it yrast} eigenstate of the balanced system containing $N_\downarrow=N_\uparrow=N=5$ particles. It turns out that, in the present case, the weak and strong interaction regimes are separated by a bifurcation of the solutions of the Bethe ansatz equations. Following the discussion presented in \cite{Shamailov2016}, we have chosen the {\it yrast} excitation corresponding to the total momentum $P=\frac{6\pi}{L}$. 

\begin{figure}[h] 	         
\includegraphics[width=1.\columnwidth]{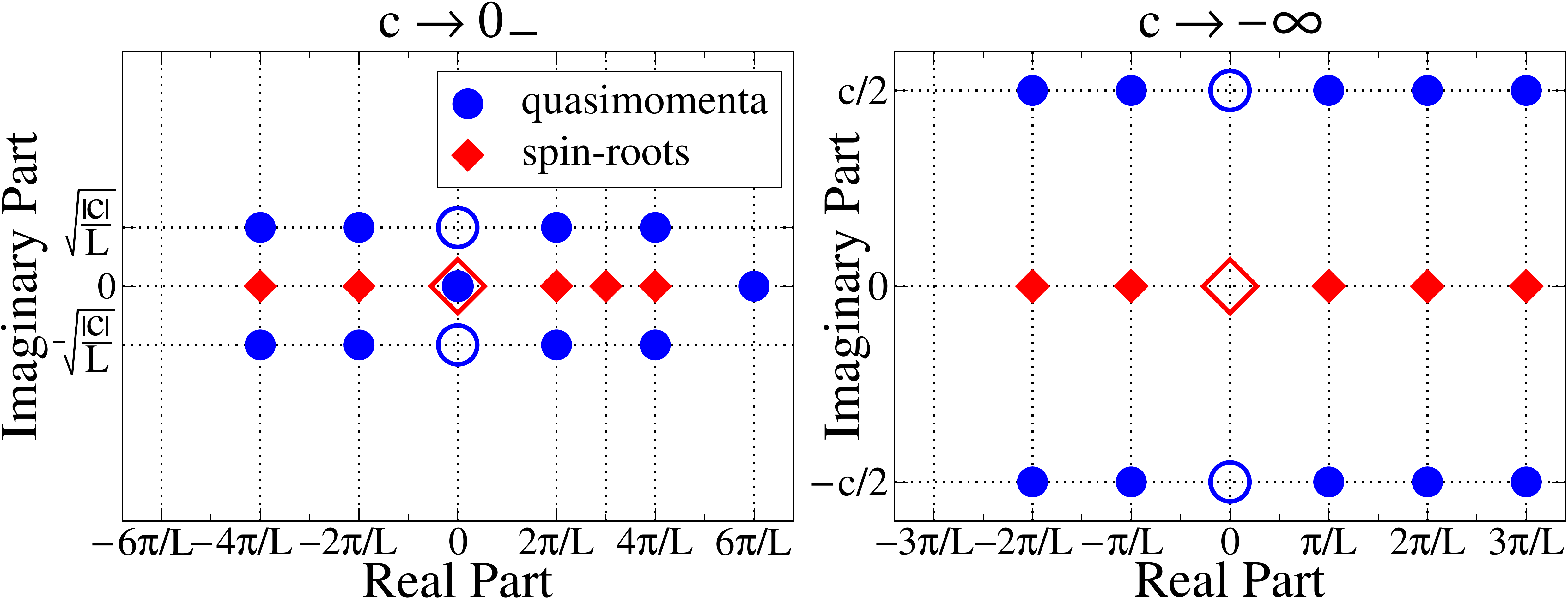}         
\caption{Scheme of the limiting solutions of the Bethe equations (\ref{GYPBCeqs1})-(\ref{GYPBCeqs2}) corresponding to the {\it yrast} state with total momentum $P=\frac{6\pi}{L}$ obtained for an unpolarized system containing $N=5$ particles in each component. Left (Right) plots in the complex plane refer to the weakly (strongly) attractive limit $c\rightarrow 0_-$($c\rightarrow -\infty$). Filled circles and diamonds show the resulting quasi-momenta and spin-roots, respectively. In comparison to the ground state solutions (see Fig.~\ref{f1}), just a few values (indicated by empty symbols) have been modified.  All units are dimensionless.
}  
\label{f5}   
\end{figure}

In the small $|c|$ limit, where the binding energy in Eq.~(\ref{GsEnMom}) changes linearly with $|c|$, one immediately notices that, in order to excite the lowest energy eigenstate belonging to the subspace of $P=\frac{6\pi}{L}$, it is energetically favorable to break the pair of quasi-momenta $k=\pm i \sqrt{|c|/L}$ from the ground state parametrization (see Fig.~\ref{f1}) and set one of them to $k=0$ and the other one to $\frac{6\pi}{L}$. Such maneuver fulfills the total momentum requirement but still it is not clear what is the structure of the corresponding spin-roots. As before, by numerical investigations and the expansion of the Bethe equations (\ref{GYPBCeqs1}), (\ref{GYPBCeqs2}) we easily find the limiting values of the quasi-momenta and the spin-roots for the {\it yrast} state in question (see the left panel of Fig.~\ref{f5}). 
Note that the scheme of the {\it yrast} excitation resembles the collective excitation of a single component Fermi gas discussed in \cite{Damski2002,Karpiuk2002a,Karpiuk2002b}. Indeed, for the very weakly interacting 5+5 particle system, the considered {\it yrast} state is represented by the following superposition 
\be
\left| \Psi \right> \approx \frac{\sqrt{2}}{2}\left( \left|\{y\}  \right>_\downarrow \left|\{g\}  \right>_\uparrow     +   \left|\{g\}  \right>_\downarrow \left|\{y\}  \right>_\uparrow \right),
\label{superposition}
\ee
where the Fock states ($\sigma=\downarrow,\uparrow$)
\be
\begin{array}{lll}
\left|\{g\}\right>_\sigma=\left|\ldots,0_{-3},1_{-2},1_{-1},1_0,1_1,1_2,0_3,\ldots\right>, \\
\left|\{y\}\right>_\sigma=\left|\ldots,0_{-3},1_{-2},1_{-1},0_0,1_1,1_2,1_3,0_4,\ldots\right>,
\end{array}
\label{superpositionFocksStates}
\ee
describe the occupation (i.e. $0_j$ or $1_j$) of single particle momentum states $\propto \mathrm{exp}[i2\pi j x/L]$. This result stays in agreement with the BCS prediction that, in the weakly interacting case, only one component of the Fermi gas has to be collectively excited in order to reproduce dark soliton features \cite{SachaDelande2014}. We stress that the BCS regime, where the Cooper pairs are much larger than the average interparticle distance, is very hard to simulate numerically within the Bethe approach. Indeed, in order to deal with such Cooper pairs the system must be much larger, i.e. the total number of particles $N\gg 5$, that is not attainable with the current computer resources.
 
 On the other hand, in  the strongly attractive limit, the binding energy increases very quickly $\sim c^2$. Hence, to deal with the {\it yrast} state one cannot break any pair of the ground state quasi-momenta. This case is closely related to the type II excitations known from the Lieb-Liniger model \cite{Korepin93,Gaudin,Lieb63,Lieb63a,sato12,sato12a,sato16,Syrwid2015,Syrwid2016}. It was also pointed out in Ref.~\cite{Shamailov2016} that, in the $c\rightarrow - \infty$ limit, the {\it yrast} excitation relies on the shift of a pair of quasi-momenta just above the Fermi surface. The same thing has to be done with the corresponding spin-root. In our case, the pair $k=\pm  i\frac{c}{2}$ and the spin-root $\Lambda=0$, related to the ground state, have to be moved to the values $k=\frac{3\pi}{L}\pm i\frac{c}{2}$ and $\Lambda=\frac{3\pi}{L}$ (see the right panel of Fig.~\ref{f5}). Such an excitation scenario is very similar to the one-hole excitation in the Lieb-Liniger model that reveals totally dark soliton structures \cite{Syrwid2015,Syrwid2016}. That is, when the momentum per boson of an {\it yrast} exictation approaches $\pm\frac{\pi}{L},$ one expects a single point in configuration space where the wave function of the last boson reveals a  phase flip by $\pi$  indicating the presence of a dark soliton density notch. Therefore, we have chosen the total momentum $\frac{6\pi}{5L}$ per dimer in the hope of observing clearly visible dark soliton signatures in the two-component Fermi gas.

\begin{figure}[h] 	         
\includegraphics[width=1.\columnwidth]{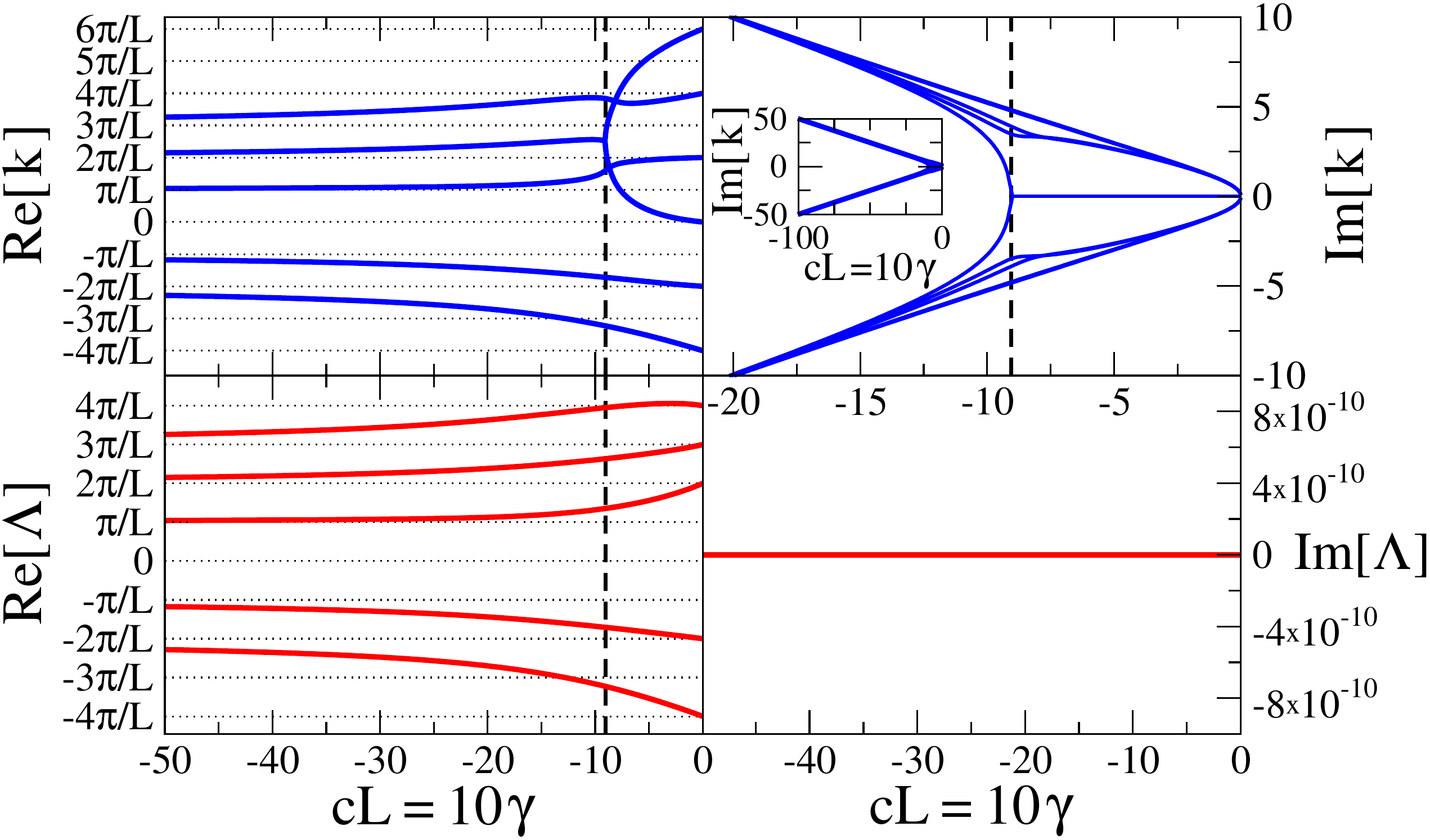}         
\caption{Quasi-momenta and spin-roots  solutions of the Bethe equations   (\ref{GYPBCeqs1}) and (\ref{GYPBCeqs2}) corresponding to the unpolarized 5+5 particles {\it yrast} state with total momentum $P=\frac{6\pi}{L}$. Left (Right) panel refers to real (imaginary) parts of $k_j$ and $\Lambda_j$  versus the coupling constant $c$. Note that the results approach to the limiting solutions shown in Fig.~\ref{f5}. Vertical dashed lines indicate the interaction strength [$cL \approx -9.05\, (\gamma \approx -0.905)$] where the bifurcation takes place. All units are dimensionless.
} 
\label{f6}
\end{figure}

The structure of the Bethe solutions for the {\it yrast} eigenstate changes dramatically between weakly and strongly attractive regimes. Following the step by step procedure \cite{Shamailov2016} mentioned in Sec. \ref{GS}, we obtained the {\it yrast} solutions of the equations (\ref{GYPBCeqs1})-(\ref{GYPBCeqs2}) for $N=5$ and $P=\frac{6\pi}{L}$ in a wide range of attraction strength (see Fig.~\ref{f6}). The resulting quasi-momenta reveal a bifurcation around $cL\approx-9.05\, (\gamma \approx -0.905),$ where we observe a transition between two different parametrization scenarios of the same {\it yrast} eigenstate. The bifurcation point coincides with the interaction strength where the pairing of fermions starts to be dominated by two-body physics. Indeed, we have justified in Sec.~\ref{GS} that the size of dimers is comparable to the mean separation between fermions of the same kind when $\gamma \approx -1$ and the pairing is dominated by two-body physics for $\gamma \lesssim -1$.  
 
\begin{figure}[h] 	         
\includegraphics[width=1.\columnwidth]{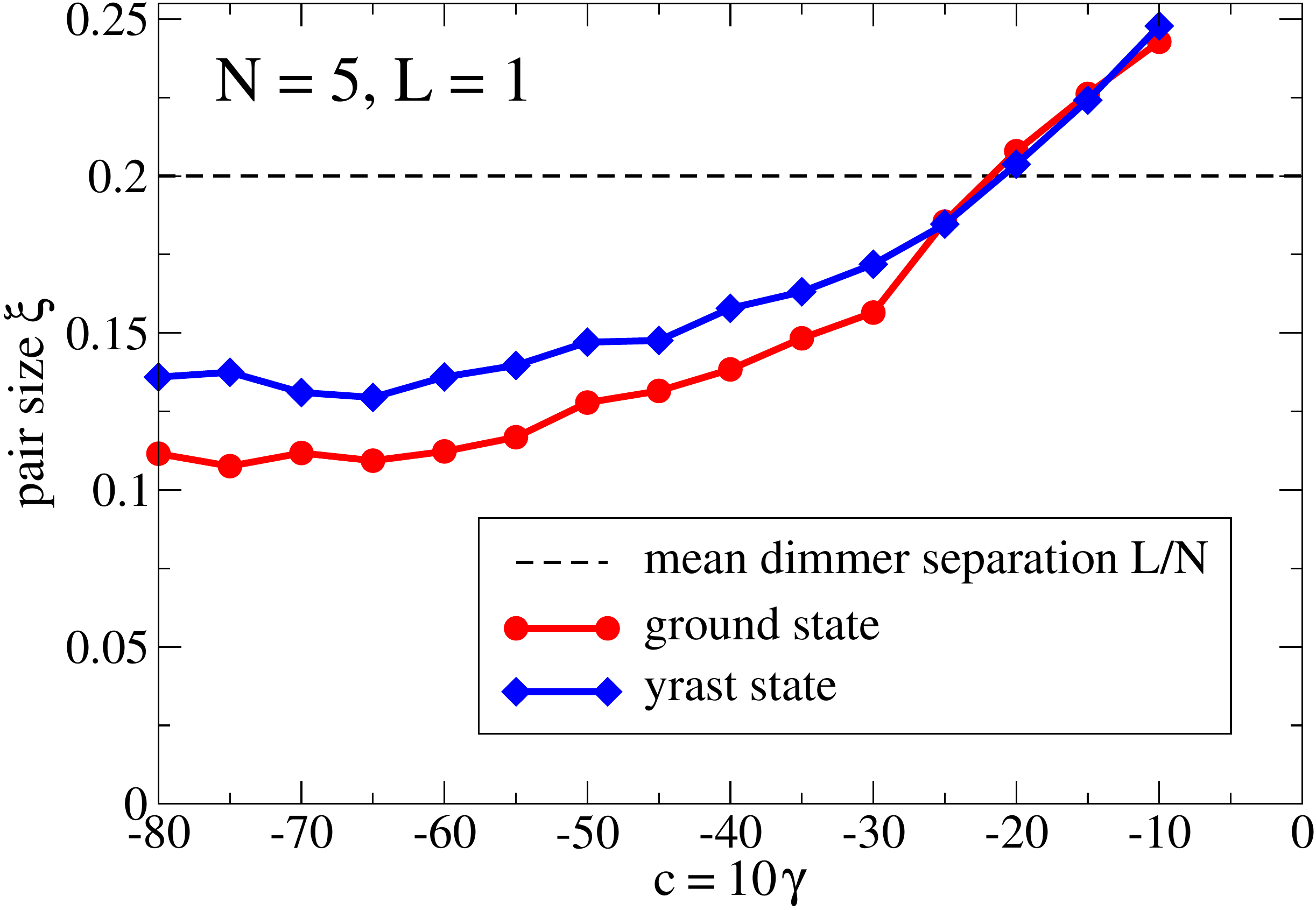}
\vspace{-0.3cm}         
\caption{Dependence of the pair size $\xi$ defined in Eq.~(\ref{PairSize}) versus the interaction strength $c=10\gamma$  in the unpolarized $N_\uparrow=N_\downarrow=N=5$ particles system of size $L=1$ prepared in the ground state (red curve with filled circles)  and in the {\it yrast} state (blue curve with filled diamonds) with $P=6\pi$. Note that both curves overlap for $c\ge-25$ ($\gamma\ge-2.5$). For stronger interactions, the pair size is slightly larger for the {\it yrast} eigenstate than for the ground state. The average distance between dimers $\bar{\delta}=L/N=0.2$ is indicated by the black dashed line. All units are dimensionless.
} 
\label{f7}
\end{figure}

We expect that tightly bound $\downarrow$--$\uparrow$ molecules appear for $\gamma \lesssim -1$. For the considered {\it yrast} state, we can carry out the approach used for the ground state, i.e. by collecting $\widetilde{\Delta}$, defined in Eq.~(\ref{PairsRelDist}), in many realizations, we can compute the distributions of relative distance between paired fermions. The dimer size is determined by the expectation value of $\widetilde{\Delta}^2$. That is, the quantity
\be
\xi=2\sqrt{\frac{1}{n}\sum_{j=1}^n\widetilde{\Delta}_{j}^2},
\label{PairSize}             
\ee
where $n$  denotes the number of collected $\widetilde{\Delta}$ distances, is related to the width of the two-body probability distribution in Eq.~(\ref{ProbDens2Body}). The quantity $\xi$ has been calculated in a wide range of interaction strength for the ground state and for the {\it yrast} state with $P=\frac{6\pi}{L}$.  It turns out that, for $c\gtrsim-25$ ($\gamma \gtrsim-2.5$), the results for both  states overlap as one can see in Fig.~\ref{f7}.  For $c\lesssim-25$ ($\gamma\lesssim -2.5$), we enter the stronger interaction regime where the pairing is definitely dominated by two-body physics. Then, the average dimer size $\xi$ is slightly larger for the {\it yrast} state than for the ground state. Such a behavior can be attributed to the fact that there is much more kinetic energy in the excited eigenstate than in the ground state.
\vspace{-0.2cm}    

\vspace{-0.1cm}     

The very strongly attractive regime is very difficult to study numerically. To satisfy the periodic boundary conditions, we have to operate with quadruple precision which turns out to be insufficient when $\gamma<-7$. Such requirements come from the fact that in the analytical expression, presented in Eq.~(\ref{GYSol}), of the wave function $\Psi,$ there are exponentials $e^{ik_jx_n^\sigma}:$ for complex quasi-momenta $\Im(k_j)\approx \pm i\frac{c}{2}$ with a large $c,$ the quadruple precision is insufficient. Moreover, in order to properly reproduce the investigated distributions by means of the Metropolis procedure, one needs more steps of a Markovian walk for strong attraction. Hence, we restrict our further studies to $\gamma \ge -7$.

\section{Yrast state: Emergence of dark soliton signatures}
\label{Yrast}
 
 {\it Yrast} states in the Yang-Gaudin model are expected to be strictly connected with dark solitons \cite{Shamailov2016}. Because of the ring geometry of the system, the eigenstates are translationally invariant. Therefore, starting with an eigenstate of translationally invariant system, the corresponding reduced single particle probability density cannot possess any soliton-like features. We expect that dark soliton structures can emerge due to the spontaneous breaking of the translational symmetry induced by measurements of positions of particles, like in  the Lieb-Liniger model \cite{Syrwid2015,Syrwid2016,Syrwid2017}. Starting with the {\it yrast} eigenstate of the  balanced $N_\downarrow=N_\uparrow=N$ system,  we have performed numerical simulations of the measurement of positions of $N_\uparrow-1$ spin-up fermions and of $N_\downarrow-1$ spin-down fermions. Then, we know the positions of $2N-2$ particles, $\widetilde{x}^{\downarrow,\uparrow}_{j=1,\ldots,N-1}$, and the wavefunction of the last  two fermions reads 
 \be
\Psi_{2}(x^{\downarrow},x^{\uparrow})=\Psi(\{\widetilde{x}^\downarrow_{1,\ldots,N-1},x^\downarrow\} ,\{ \widetilde{x}^\uparrow_{1,\ldots,N-1},x^\uparrow\}).
\label{2D_WF}             
\ee 
In the following we analyze properties of the above Eq.~(\ref{2D_WF}).  
We consider two different ways of measuring the positions of $2N-2$ particles: 
\begin{enumerate}
\item We can assume that the measurement takes place when two fermions with spin up and down are detected at the same positions, i.e. $x^\downarrow_s =
x^\uparrow_s$ for $s = 1, 2, \ldots, N-1$.  It resembles the rapid ramp technique used in experiments where a sweep across a Feshbach resonance leads to the creation of tightly bound molecules \cite{Mies2000,Abeelen1999,Yurovsky1999,Regal2003,Cubizolles2003,Jochim2003,Strecker2003,Diener2004,Perali2005,Altman2005,Yuzbashyan2005}  and a subsequent measurement of the molecular density is performed. We will call this measurement "zero-size" because in this case the size of pairs of $\downarrow$--$\uparrow$ fermions is zero.

\item We can perform the measurement of positions of particles by sampling the many-body probability density without any constraint. In other words, we do not measure the pairs but single particles. This kind of the measurement will be dubbed  "any-size"  because we assume that pairs of fermions can have any size.
\end{enumerate} 

\begin{figure}[t!] 	         
\includegraphics[scale =0.2253 
]{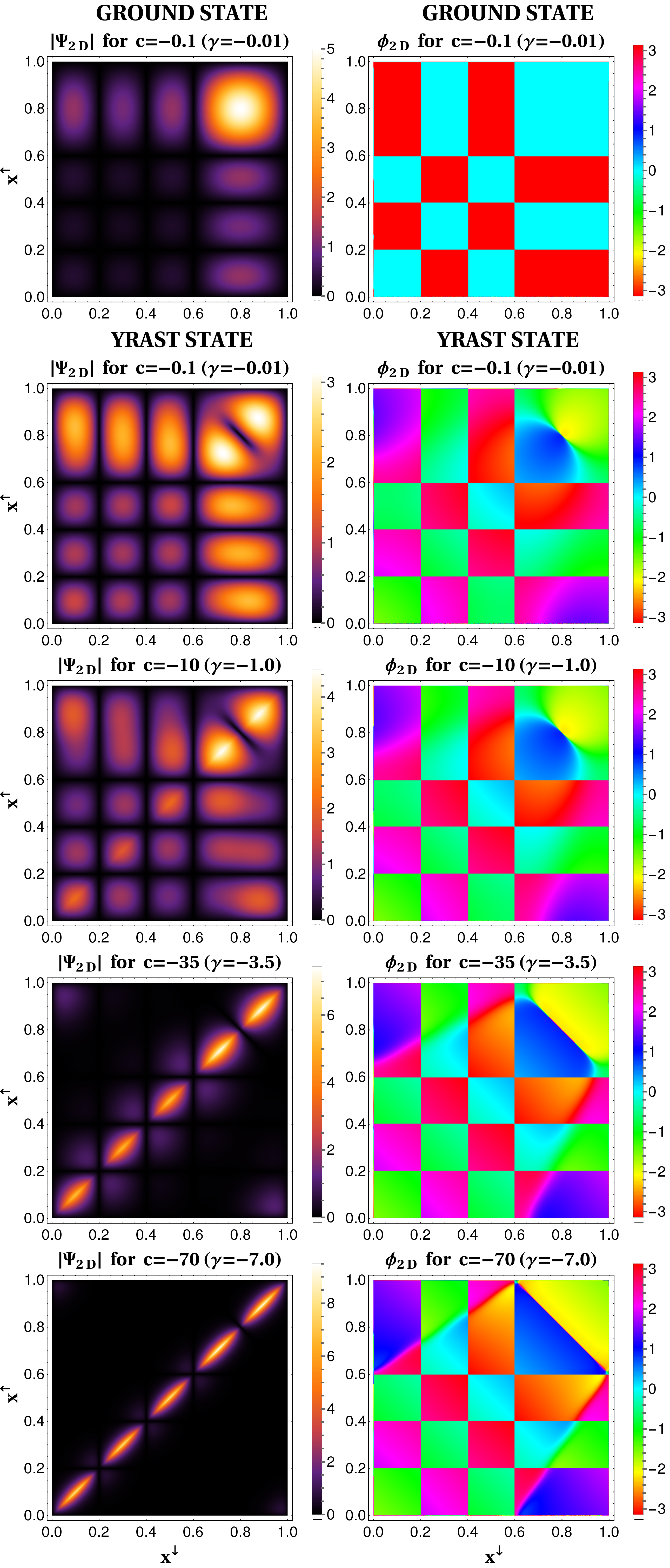}  
\vspace{-0.255cm}   
\caption{Wave function $\Psi_{2}(x^\downarrow,x^\uparrow)=|\Psi_{2}|\mathrm{e}^{i \phi_{2}}$ for the last two fermions with opposite spins in the 5+5 particles system on a ring of length $L=1$. We consider the {\it yrast} state corresponding to total momentum $P=6\pi$ for different coupling strengths, i.e. $-70\le c \le -0.1$ $(-7\le \gamma \le -0.01)$. It is assumed that four pairs of spin up and down fermions (of zero size) have been measured at the positions corresponding to the maximal probability density, e.g. $\widetilde{x}_j^\downarrow=\widetilde{x}_j^\uparrow =\frac{j-1}{5}$ for $j=1,2,3,4$.  The resulting amplitudes (left panels) and phase (right panels) of the wave function of the last two fermions show a density notch and a phase flip along the diagonal, respectively. Such dark soliton-like signatures appear around $x^\downarrow\approx x^\uparrow \approx 0.8$, i.e. exactly between most distant dimers that have been measured. For comparison, the ground state wave function $\Psi_{2}(x^\downarrow,x^\uparrow)$, which can be chosen as a real-valued function, is depicted in the first row. All units are dimensionless.
\vspace{-0.855cm}   
}  
\label{f8}    
\end{figure}

Let us start with an analysis of a single realization of the zero-size measurement process.  We consider an unpolarized system of $N_\downarrow=N_\uparrow = N =5$ particles, $L = 1$ and a wide range of the interaction strength, i.e. $-70\le c \le -0.1$ $(-7\le \gamma \le -0.01)$. The system is prepared initially in the {\it yrast} state with total momentum $P=6\pi$ which is analyzed in  Sec.~\ref{YrastPair}. The structures of the observed two-body wave function defined in Eq.~(\ref{2D_WF}) 
depend on the positions where the first $2N-2$ particles are measured. 
For a small particle number, a single realization of the measurement process may result in a two-particle wave function in Eq.~(\ref{2D_WF}) which does not clearly show dark soliton signatures. Therefore, we have decided to choose optimal positions of the measured pairs of fermions, that is a configuration of the pairs that corresponds to the maximal value of the probability density, 
e.g. $\widetilde{x}_j^\downarrow=\widetilde{x}_j^\uparrow =\frac{1}{5}(j-1)$ for $j=1,2,3,4$.  This choice means that due to the Pauli exclusion rule the last remaining pair of fermions is likely to be detected in  the largest free space interval, namely between $x^\downarrow=x^\uparrow=0.6$ and 1.   The phase flip is expected to be observed around $x=0.8$ because it minimizes the energy. The modulus and phase of the resulting two-body wave functions $\Psi_{2}(x^\downarrow,x^\uparrow)=|\Psi_{2}|\mathrm{e}^{i \phi_{2}}$, see Eq.~(\ref{2D_WF}), are presented in Fig.~\ref{f8}. It turns out that, in such a case, the  amplitudes $|\Psi_{2}|$ and phase distributions $\phi_{2}$ reveal clearly visible a density notch and a phase jump localized around expected position $x^\downarrow\approx x^\uparrow \approx 0.8$  both for weak and strong attraction. Such dark soliton signatures	 do not emerge in the course of the particle detection process when we start with the ground state (for comparison, see top panels of Fig.~\ref{f8}). Note that, as expected, the stronger interactions we deal with, the more dominant the diagonal elements of the density are. Furthermore, thanks to the Pauli exclusion principle, in the plots of the phase distributions, one notices a nodal structure at the positions of the initially measured zero-sized pairs of $\downarrow$--$\uparrow$ fermions. Such nodal structures are present for any interaction strength, both for the ground state and the {\it yrast} eigenstate and can be explained by a simple reasoning. That is, the wave function describing two identical noninteracting fermions can be cast into the form
 \be
\varphi(x,x+\varepsilon)\propto \mathrm{e}^{i\alpha x}\mathrm{e}^{i\frac{\alpha}{2} \varepsilon}\sin\left(\beta\varepsilon\right), \,\,\,\,\,\,\, \alpha\in \mathbb{R}, \,\, \beta \in \mathbb{R}_+ , 
\label{2D_WF_FreeFermions}             
\ee 
that simply reveals the $\pi$-phase flip at $x$ when $\varepsilon$ passes through zero.   

\begin{figure}[h] 	         
\includegraphics[width=1.\columnwidth]{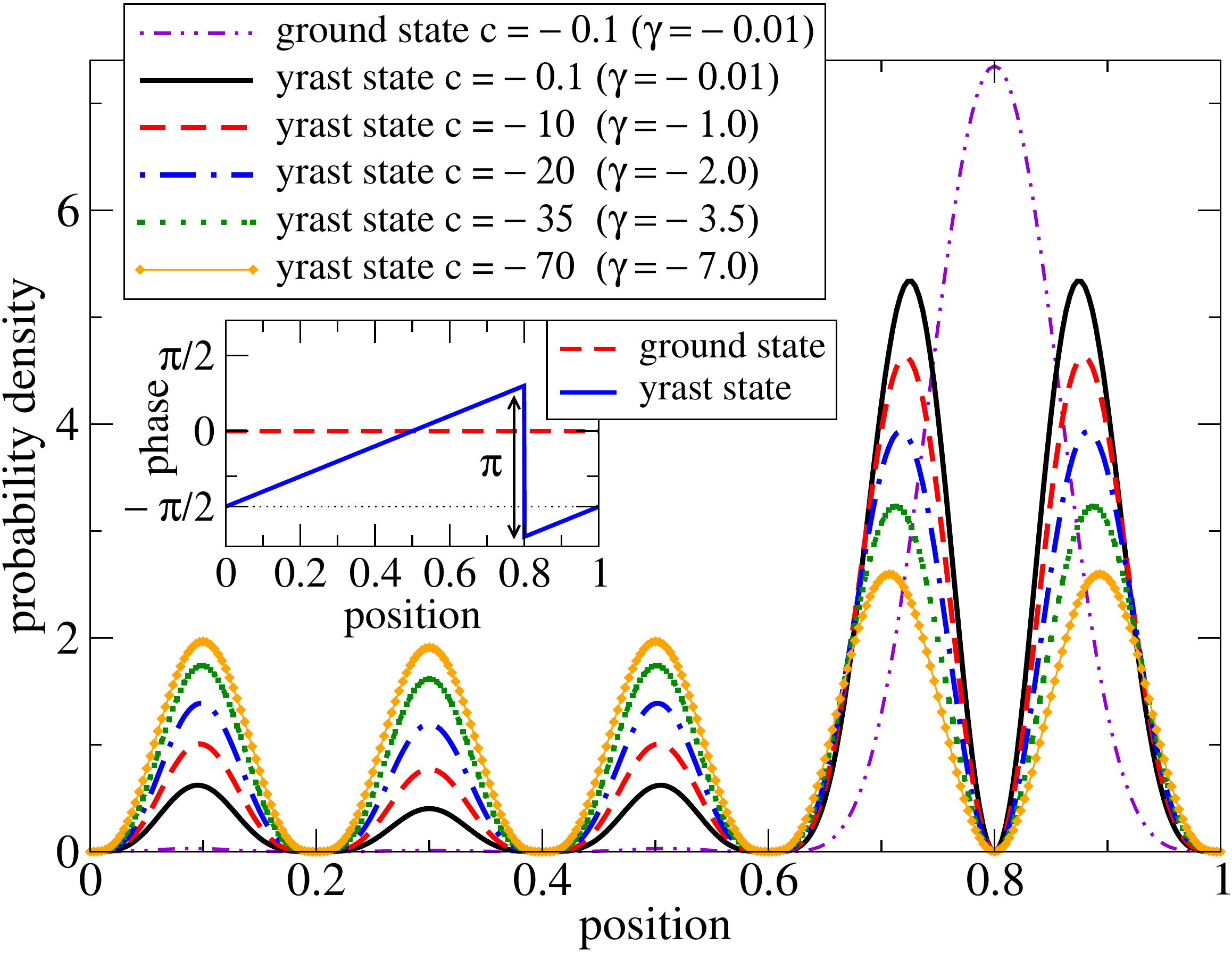}         
\caption{Diagonal part of $|\Psi_{2}(x^\downarrow,x^\uparrow)|^2$ shown in Fig.~\ref{f8}. The inset presents the corresponding diagonal phase distribution $\phi_{2}$ which is identical for all coupling constants $c<0$. The density notch is clearly visible around $x=0.8$ and its position coincides with the position of the phase flip. For comparison, the same numerical experiment was performed for the ground state for weak attraction $c=-0.1$ $(\gamma=-0.01)$ when no soliton-like structure can be observed around $x=0.8$. All units are dimensionless.
}
\label{f9}  
\end{figure}

Figure~\ref{f9} are cuts along the diagonals of the two-dimensional plots shown in Fig.~\ref{f8}, i.e. we present the probability density and the phase of $\Psi_{2}(x,x)$.
We observe dark soliton signatures like density notch and phase flip at $x=0.8,$ and also notice that the distance between the two main peaks of probability density around $x=0.8$ slightly increases  when $\gamma$ becomes more negative. Note that the {\it yrast} state for very weak interactions corresponds to a single fermion excitation (see Eqs. (\ref{superposition}) and (\ref{superpositionFocksStates}) as well as left panel of Fig.~\ref{f5}), while, in the strong attraction limit, it is related to the excitation of a single pair of fermions (cf. right panel of Fig.~\ref{f5}). In order to keep the same total momentum $P$, the momentum of a single fermion in the former case has to be twice larger than the momentum of each fermion in the latter case, cf. Fig.~\ref{f5}. Consequently, we deal with longer wavelengths for strong interaction which can be responsible for the observed increase of the distance between the two peaks around the density notch. In other words, the shorter wavelengths we deal with, the narrower structures can be reproduced in the density. Since one deals with a Fermi system, the wave function $\Psi_{2}(x,x)$ vanishes at the positions of the initially measured dimers. For comparison, we also depict the results for the ground state in the weakly interacting case, for $\gamma=-0.01$. One immediately notices that no soliton signature around $x=0.8$ can be observed in this case.

So far we have analyzed dark soliton signatures in the wave function of the last pair of fermions when the $N-1$ pairs of zero-size are assumed to be measured at the equidistant positions. Such a configuration allows us to observe clearly the dark soliton signatures. However, in experiments pairs of fermions or fermions themselves are detected at random positions according to the probability density of the {\it yrast} state. Now, preparing the system initially in the same {\it yrast} state as before, we investigate the diagonal probability density and the phase distribution for the last pair of  $\downarrow$--$\uparrow$  fermions averaged over many realizations of the particle detection process. By employing the Metropolis routine and the Bethe ansatz approach, we have performed numerical simulations of the measurement of $N-1=4$ dimers assuming that they have either zero-size or any-size (i.e. measurement of particles without any additional restrictions).  By the fact that we deal with periodic boundary conditions the position of the phase flip indicating the soliton-like structure varies randomly from one measurement realization to another one. 
In order to determine average distributions we shift all the results so that the corresponding phase flip is always located at $\frac{L}{2}=0.5$. In every single realization of the detection process, the diagonal phase distribution $\phi_2(x,x)$ of the last $\downarrow$--$\uparrow$ pair reveals a flip. To satisfy the periodic boundary conditions the relation $\phi_2(L,L)-\phi_2(0,0)=2\pi J$, where in general $J\in\mathbb{Z}$, has to be fulfilled. In the limiting case where the soliton is completely dark (i.e. when the density drops to zero like in Fig.~\ref{f9}), the phase flip occurs abruptly at a single point, i.e. $\lim_{\epsilon\rightarrow 0}\left[\phi_2(x_S\!+\!\epsilon,x_S\!+\!\epsilon)-\phi_2(x_S\!-\!\epsilon,x_S\!-\!\epsilon)\right]=\pm\pi$ modulo $2\pi$, where $x_S$ is the soliton position. Then, all $J\in \mathbb{Z}$ become equivalent and cannot be distinguished. Therefore, in such a case we show the phase plot corresponding to $J=0$ only, see the instet in Fig.~\ref{f9}.

\begin{figure}[h] 	         
\includegraphics[width=1.\columnwidth]{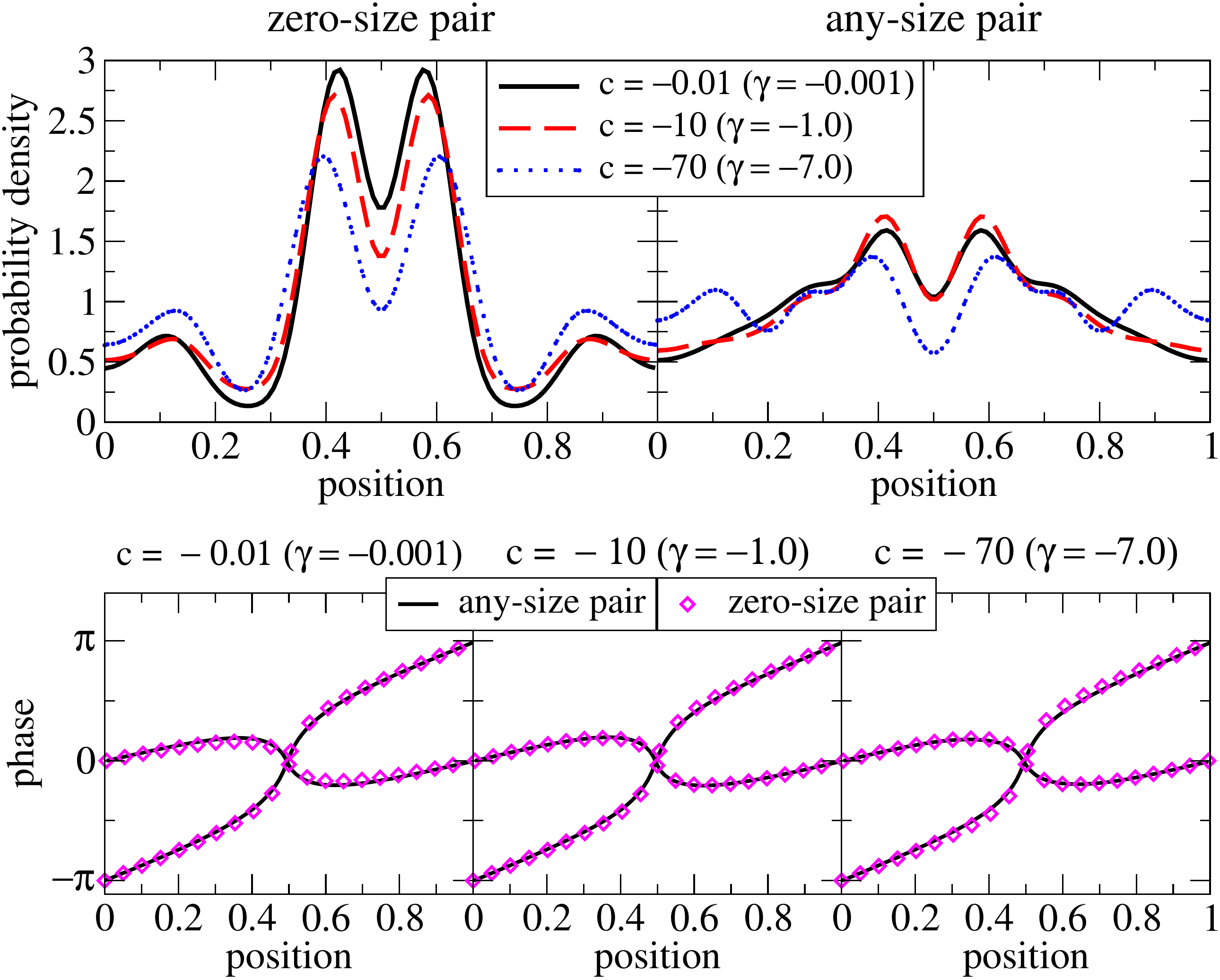}          
\caption{Diagonal probability densities and phase distributions for the last pair of $\downarrow$--$\uparrow$ fermions averaged over many (about ten millions) realizations of the detection process for a system of $N_\downarrow=N_\uparrow=5$ fermions prepared in the {\it yrast} state with total momentum $P=6\pi$ (the system size $L=1$). Upper panels show the averaged densities for different attraction strengths obtained using two alternative initial measurement schemes: left (right) panel corresponds to zero-size (any-size) of fermionic pairs. Lower panels display the averaged phases obtained within different detection schemes and for different attraction strengths. Note that they are almost identical independently of $\gamma$ and of the applied detection scheme. They reveal down ($J=0$) and up ($J=1$) phase flips. All units are dimensionless.
} 
\label{f10}
\end{figure}        

The results presented in the upper panels of Fig.~\ref{f10} show the density notches in 
the diagonal probability density for the last fermions $|\Psi_2(x,x)|^2$ defined as in Eq.~(\ref{2D_WF}), averaged over many realizations of the measurement of $2N-2$ fermions. The notches are clearly visible for all interaction strengths independently of the measurement scheme (zero-size or any-size) of the detected fermionic pairs. Moreover, we always observe phase flips of two types: facing up ($J=1$) and down ($J=0$) that are collected separately in lower panels of Fig.~\ref{f10}. In contrast to the case of the equidistant measurement of zero-size dimers -- cf. Figs.~\ref{f8}-\ref{f9} -- the average density notches do not drop to zero, i.e. the soliton structure is not completely dark. 
The shape of the average density notch  depends on the measurement scheme: the density notch is sharper with the zero-size scheme than for the any-size one. Like in Fig.~\ref{f9}, the distance between the two main peaks around the density notch increases with the interaction strength.
This is related to a decrease of accessible momenta in the {\it yrast} state. Suprisingly, 
the measurement scheme and the interaction strength almost do not affect the shape of the average phase distribution.

\begin{figure}[h] 	         
\includegraphics[width=1.\columnwidth]{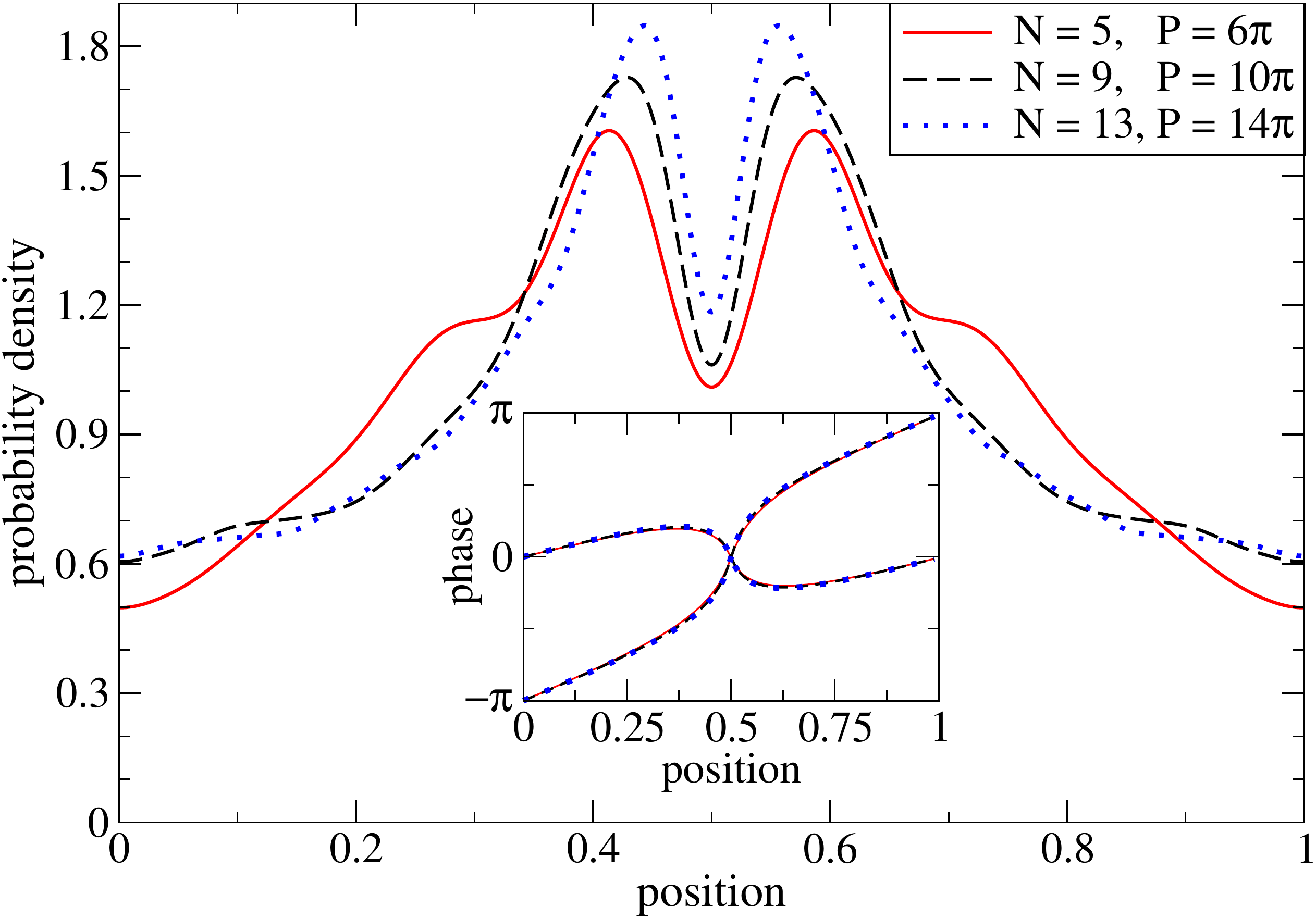}         
\caption{Average probability density for the last zero-size pair of $\downarrow$--$\uparrow$ fermions for different initial number of particles $N=N_\downarrow=N_\uparrow=5$, 9, 13. The system of size $L=1$ was initially prepared in the {\it yrast} state with total momentum $P=\pi(N+1)$. The more particles in the system, the narrower the density notch visible at the center of the plot.
	The inset displays the corresponding (nearly identical for all $N$ considered here) average phase distributions of two kinds: facing up (with $J=1$) and down ($J=0$).  All the calculations were performed in the presence of weak attraction $\gamma=-0.01$.  All units are dimensionless.
} 
\label{f11}
\end{figure}         
 
The last thing we would like to consider is the influence of the particle number $N=N_\downarrow=N_\uparrow$ on the soliton structures. For weak interactions, systems containing more than $N=5$ particles in each component can be studied by numerical diagonalization of the Hamiltonian in Eq.~(\ref{h}) (see Sec. \ref{NumMet}). In order to compare $N>5$ particles systems  with the $N=5$ results explored so far in this paper, we have to choose exactly the same type of an {\it yrast} excitation. For this purpose, we investigate odd numbers of particles in each component up to  $N=13$. The {\it yrast} excitation corresponds to the total momentum $P=\pi (N+1)$ for the system of size $L=1$. Converged results can be obtained only for weak attraction. By comparing with results obtained using the Bethe ansatz, we found that current computer resources allows us to study $N=5$ systems via numerical diagonalization up to $\gamma \gtrsim -0.2$ only. By investigating the wave function for the last pair of $\downarrow$--$\uparrow$ fermions we, in fact, calculate higher order correlation functions. If so, the eigenstate corresponding to the {\it yrast} state has to be determined very accurately to avoid significant numerical errors in the final results. The number of basis states that have to be taken into account for fixed $\gamma$, proliferates dramatically with the increase of $N$. Such demand eliminates the possibility of analysis for interactions  stronger than $\gamma\approx -0.2$. Therefore, we restrict ourselves to $\gamma=-0.01$. We have decided to apply only the any-size measurement scheme because, for weak interactions, the average size of a $\downarrow$--$\uparrow$ pair is larger than the mean interparticle separation. In the weak coupling regime,  in order to create the {\it yrast} excitation, we need to break the pair of the quasi-momenta with zero real part and translate one of those quasi-momenta just above the Fermi surface. The Fermi momentum increases with $N$, hence, the {\it yrast} excitation requires the "injection" of a larger momentum if there are more particles in the system. The fact that with an increase of $N$ we deal with  a larger momentum "injection" and thus with shorter wavelengths is consistent with the results presented in Fig.~\ref{f11} where the width of the density notch decreases with $N$. 
We also observe the phase flips with winding numbers $J=0$ and $J=1$.

\section{Conclusions}
\label{conclusions}

We have considered a one-dimensional two-component gas of ultracold fermions interacting via an attractive Dirac-delta potential with periodic boundary conditions. The system is described by the Yang-Gaudin Hamiltonian, see Eq.~(\ref{h}), that can be solved analytically with the help of the Bethe ansatz. Since the Hamiltonian is invariant under spatial translations of all particles, one cannot observe any feature of the eigenstates by looking at the reduced single particle density only. Therefore, by employing the Metropolis algorithm, we performed numerical simulations of the measurement of positions of particles. Starting with the unpolarized 5+5 particles system in the ground state, we investigated the formation of dimers in a wide range of the attraction strength. The analysis showed that the average size of pairs of $\downarrow$--$\uparrow$ fermions becomes smaller than the mean dimer separation $\bar{\delta}$ when the effective dimensionless interaction parameter $\gamma\lesssim-1$. In such a regime, the many-body distribution of the relative distance between fermions  follows the two-body prediction given by Eq.~(\ref{ProbDens2Body}). When the attraction between fermions with opposite spins is weak (i.e. when $\gamma\gtrsim-1$), one enters the regime where the size of the Cooper-like pairs is larger than $\bar{\delta}$.

The key element of this paper is an analysis of {\it yrast} eigenstates in the context of the anticipated emergence of dark soliton signatures. For this purpose, we studied a particular {\it yrast} state with total momentum $P=\pi(N+1)/L$. By successive particle detections, we analyzed the wave function for the last pair of fermions. Here, we decided to examine two different schemes of initial $N-1$ pairs measurements: either detection of zero-size dimers or without restriction on measured positions of spin up and down fermions. The results clearly show the dark soliton signatures (density notches and phase flips) in the wave function of the last remaining pair of fermions, for all interaction strengths and for both detection schemes. However, the choice of the detection schemes has an influence on the shape of the average probability densities. Surprisingly, the interaction strength and the detection schemes almost does not affect the shape of the average phase distributions, revealing a clearly visible  phase flip.  
Such a resistance of the phase flip to parameter changes resembles the behavior observed for a Bose gas described by the Lieb-Liniger model \cite{Syrwid2015,Syrwid2016,Syrwid2017}.

\section*{Acknowledgments}
Support of the National Science Centre, Poland via projects No.2016/21/B/ST2/01095 (A.S.) and No. 2015/19/B/ST2/01028 (K.S.) is acknowledged.

\end{document}